\documentclass[sigconf]{acmart}

\usepackage{booktabs} 
\usepackage{algorithm2e}
\usepackage{todonotes}
\usepackage{subcaption}
\usepackage{cleveref}
\setcopyright{none}

\acmConference[MLMH'18]{ACM KDD Workshop on Machine Learning for Medicine \& Health}{August 2018}{London UK}
\acmYear{2018}
\copyrightyear{2018}

\begin{document}
\title{Online Heart Rate Prediction using Acceleration from a Wrist Worn Wearable}

\author{Ryan McConville}
\affiliation{%
  \institution{University of Bristol}
  \city{Bristol, UK}
}
\email{ryan.mcconville@bristol.ac.uk}

\author{Gareth Archer}
\affiliation{%
  \institution{Sheffield Teaching Hospitals NHS Foundation Trust}
  \city{Sheffield, UK}
}
\email{garetharcher@nhs.net}

\author{Ian Craddock}
\affiliation{%
  \institution{University of Bristol}
  \city{Bristol, UK}
}
\email{ian.craddock@bristol.ac.uk}

\author{Herman ter Horst}
\affiliation{%
  \institution{Phillips}
  \city{Netherlands}
}
\email{herman.ter.horst@philips.com}

\author{Robert Piechocki}
\affiliation{%
  \institution{University of Bristol}
  \city{Bristol, UK}
}
\email{r.j.piechocki@bristol.ac.uk}

\author{James Pope}
\affiliation{%
  \institution{University of Bristol}
  \city{Bristol, UK}
}
\email{james.pope@bristol.ac.uk}

\author{Raul Santos-Rodriguez}
\affiliation{%
  \institution{University of Bristol}
  \city{Bristol, UK}
}
\email{enrsr@bristol.ac.uk}

\renewcommand{\shortauthors}{R. McConville et al.}

\begin{abstract}
In this paper we study the prediction of heart rate from acceleration using a wrist worn wearable. Although existing
photoplethysmography (PPG) heart rate sensors provide reliable measurements, they use considerably more energy than accelerometers and have a major impact on battery life of wearable devices. 
By using energy-efficient accelerometers to predict heart rate, significant energy savings can be made.
Further, we are interested in understanding patient recovery after a heart rate intervention, where we expect a variation in heart rate  over time. Therefore, we propose an online approach to tackle the \textit{concept drift} as time passes. We evaluate the methods on approximately 4 weeks of free living data from three patients over a number of months. We show that our approach can achieve good predictive performance (e.g., 2.89 Mean Absolute Error) while using the PPG heart rate sensor infrequently (e.g., 20.25\% of the samples).

\end{abstract}


\maketitle

\section{Introduction}    
    The importance of healthcare technology continues to grow rapidly in both consumer and clinical applications~\cite{bridging-e-health}.
    Consumers are increasingly desiring the ability to monitor and understand their health and wellbeing through the use of technology. 
    Smart and health watches are both popular devices used to do so. Clinicians are increasingly using these devices to understand patient behaviour over longer periods of time, rather than single measurements in a clinical setting.
    
    Health watches typically provide measurements of health in the form of activity detection, localisation, respiration or heart rate. The use of wrist worn devices for activity detection, such as via step counting, is widespread due to the low cost and power usage of accelerometers. 

While accelerometers have low energy usage, common heart rate measuring sensors consume significantly more power. One of the most typical approaches is photoplethysmography (PPG), which measures heart rate from the blood volume pulse seen in microvascular tissue~\cite{0031-9155-19-3-003} via a reflected light. Consumer PPG sensors typically consume up to 5000 times the power than the accelerometer~\cite{7380a8563ce24e06b8b2bc5a17155a67} used in wearables, which is an impediment to the long battery life desired in wearable technology. As accelerometers are widespread and exist in any device which would likely also contain a heart rate sensor, we are interested in considering the feasibility of acceleration as a means of predicting heart rate. We note the inherent difficult in this; namely there are many known factors which influence heart rate but will not be captured by acceleration. Further we expect the same acceleration patterns relationship with heart rate will change over time, e.g., due aging or change in level of fitness. Thus, we believe that an approach which does not \textit{adapt} to these changes will not be able to predict heart rate well from acceleration. 
      
 \section{Scenario}

The \texttt{EurValve}\footnote{\url{http://www.eurvalve.eu/}} project is building a Decision Support System (DSS) for aortic and mitral valve diseases. One strand of this involves the pervasive collection of environmental data for behavioral analysis of patients in their home~\cite{db0f1a95d9f043819baf852736e5691a,41bcc3e72fc14c7c91a5115b7681e744}, once before and two times after a heart valve intervention. 
 As part of this 42 patients have been recruited and are currently undergoing the phased deployment of a custom Smart Home in a Box (SHiB)~\cite{db0f1a95d9f043819baf852736e5691a} in their home.
 The SHiB contains four gateways which receive accelerometer data from the wrist worn wearable. 
 These gateways are then used for indoor localization~\cite{wfiot} but may also be used for local data processing and machine learning.
  This wearable contains only an accelerometer with no heart rate sensor, but has a battery life of approximately three weeks.
 Recruited patients also wear a Phillips Health watch~\cite{Hendrikx2017ClinicalEO} which is capable of measuring heart rate using PPG, but with a battery life of up to 4 days.
 For the duration of their involvement in the study patients will be wearing both devices and thus we will collect up to 6 weeks of free living data per patient.
 This presents a unique first opportunity to investigate the potential of using acceleration to predict heart rate in real-world free living conditions over significant periods of time at different recovery stages.
 
 \subsection{Desiderata}
 We will first produce a baseline approach where we train a regression model on features extracted from acceleration, aligned with the heart rate and then attempt to predict future heart rates from the acceleration alone.
 We believe this this approach will not capture the relationship between the acceleration and heart rate, due to the aforementioned challenges.
 To improve upon this, we will then propose an active learning approach that can run on the SHiB gateways, which have bidirectional communication with the wearable. This approach will predict heart rate from the streaming accelerometer data in an online fashion and be able to request the measurement of true heart rate via PPG when required. 
 
 \section{Experiment}
 The experimental evaluation will involve 3 patients who wore the device shortly after their heart valve intervention for approximately 2 weeks, and then a further 2 weeks around 2 months later.
 We will refer to the separate occasions as phases.
 The acceleration features were extracted over a one second window and are as follows:  the min, max, std, median, mean, 25th and 75th percentile, interquartile, skewness, kurtosis and number of zero crossings of each  of the three axes. The spectral energy and spectral entropy of each axis was also extracted.
  As the heart rate is recorded at one minute intervals by the Phillips Health watch, the accelerometer features are then averaged over a one minute interval to time-align with the heart rate.

 \subsection{Offline Baseline}
The baseline approach will assume the collection of all accelerometer data and heart rate data over two separate approx. 14 day monitoring phases. The objective will be, when given one minute of acceleration data, to predict the corresponding heart rate for that time point.
 In the first experiment 60\% of a the data for one patient in one phase will be used to build a model, and the remaining 40\% used for testing. We treat the data as i.i.d in this case.
 
\begin{figure}
  \centering
   \subcaptionbox{Random Forest Regressor (MAE))\label{fig:offline-results-same-mae}}{%
 	\includegraphics[width=.45\columnwidth]{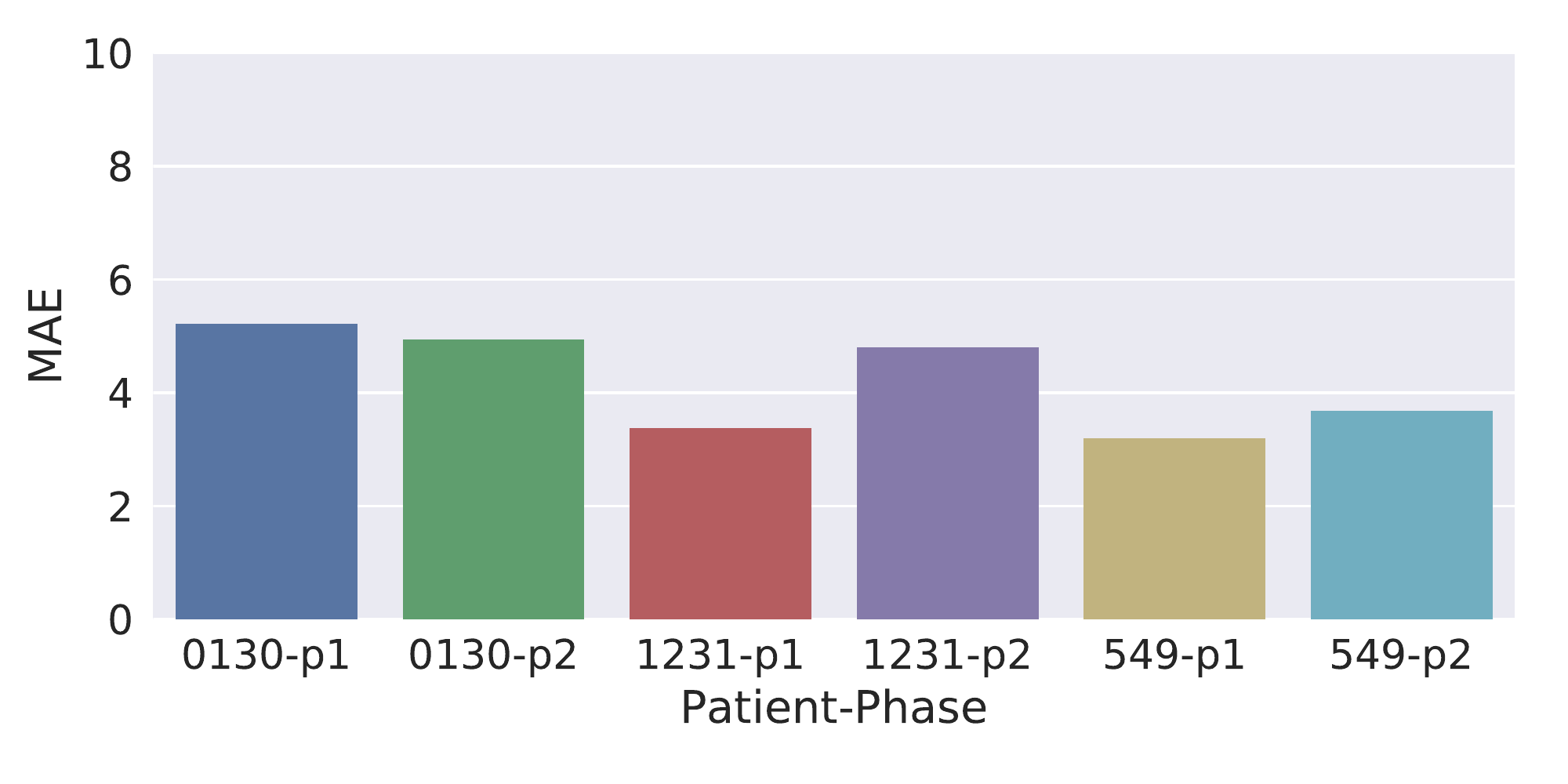} %
    }
  \subcaptionbox{Dummy Mean Predictor (MAE)\label{fig:offline-results-same-mae-dummy}}{
  	\includegraphics[width=.45\columnwidth]{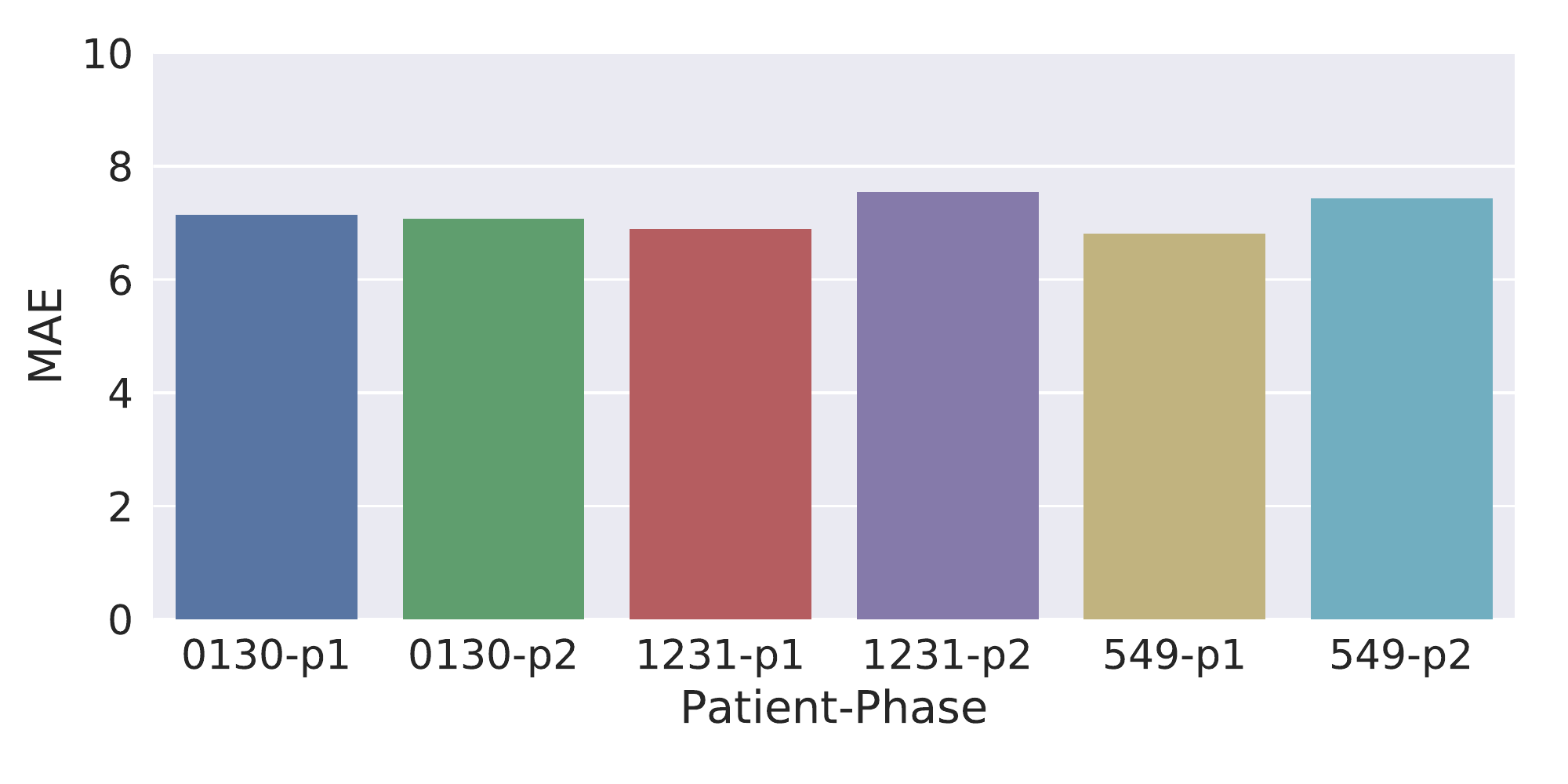}
  }
     \subcaptionbox{Random Forest Regressor (MSE)\label{fig:offline-results-same-mse}}{%
 	\includegraphics[width=.45\columnwidth]{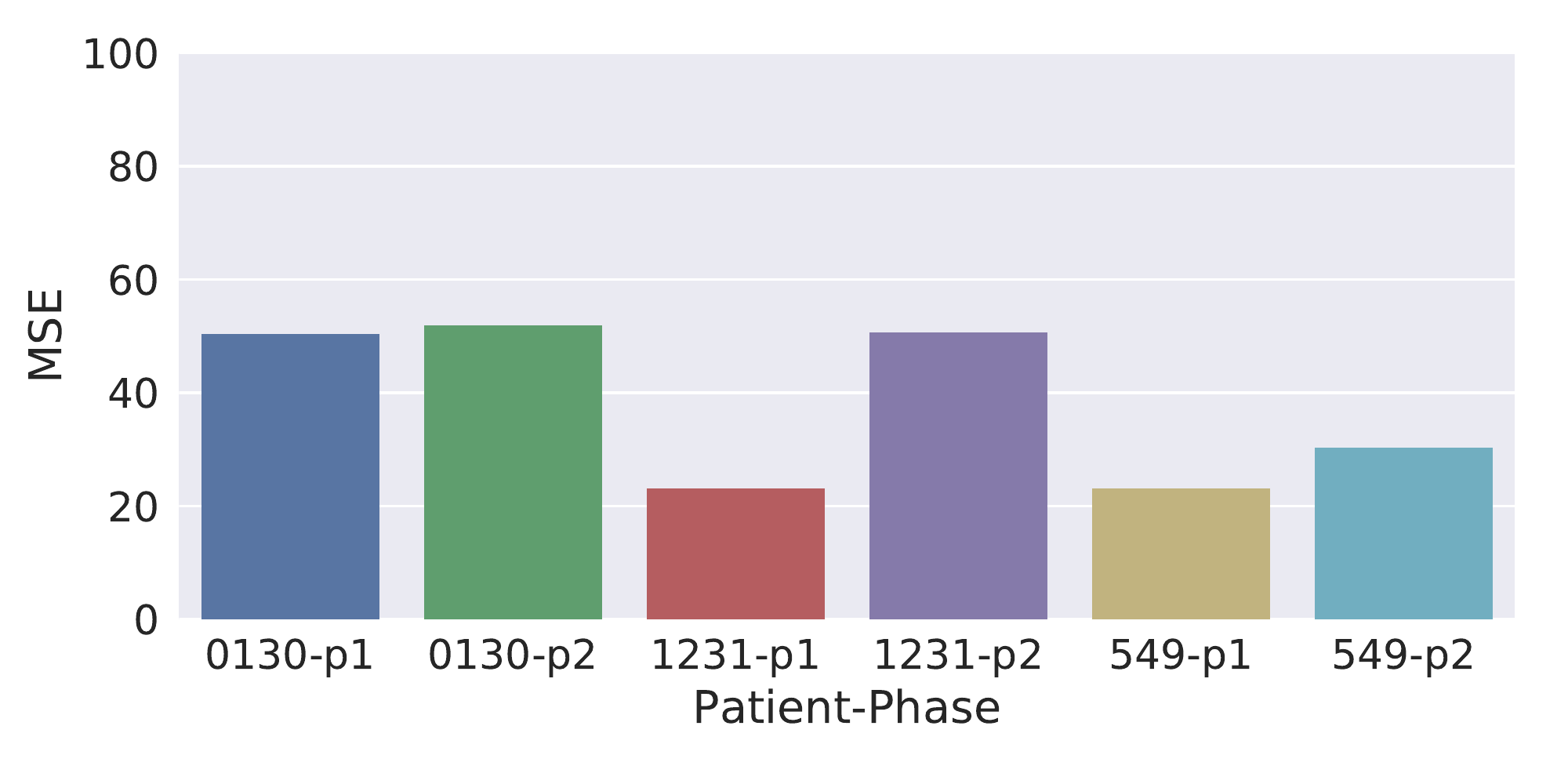} %
    }
  \subcaptionbox{Dummy Mean Predictor (MSE)\label{fig:offline-results-same-mse-dummy}}{
  	\includegraphics[width=.45\columnwidth]{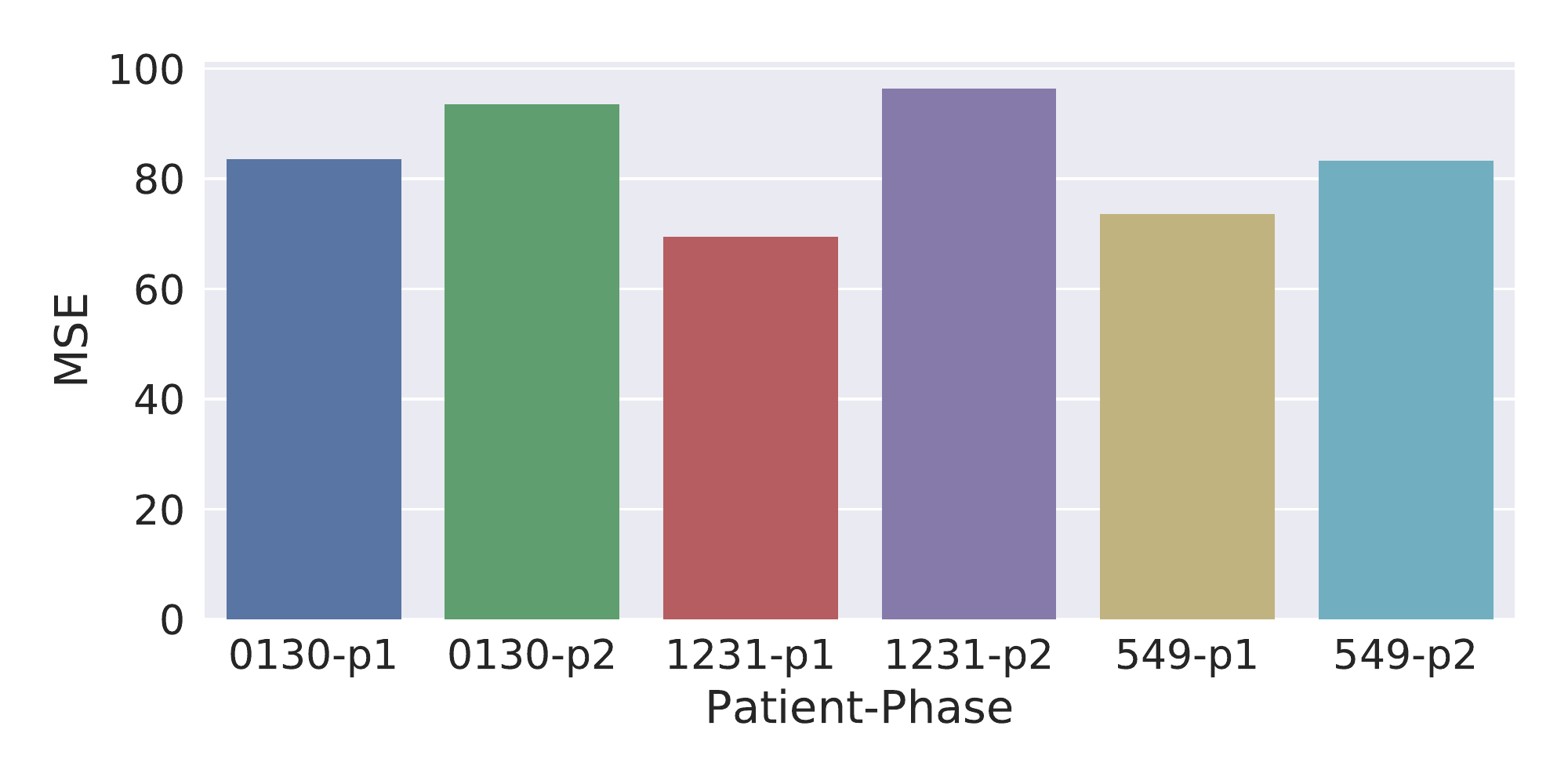}
  }
 
  \caption{MAE when tested within the same continuous approx 14 day period (a) MAE is between 4 and 6, performing better than the dummy predictor in (b). The trend is the same for MSE (c), however notably, the performance is clearly poor.}
\label{fig:offline-results-same}
  \end{figure}

Fig.~\ref{fig:offline-results-same} shows the results of this experiment. Fig.~\ref{fig:offline-results-same-mae} and \ref{fig:offline-results-same-mse} demonstrate reasonable performance is achieved with an Mean Absolute Error (MAE) between 4 and 6 and Mean Squared Error (MSE) between 23 and 51 depending on the patient and phase that was evaluated. Fig.~\ref{fig:offline-results-same-mae-dummy} shows the results of a `dummy' predictor which simply outputs the mean heart rate at each minute. This can be up to 2x worse than the acceleration based model.

\begin{figure}
  \centering
   \subcaptionbox{MAE\label{second-subfig}}{%
 	\includegraphics[width=.48\columnwidth]{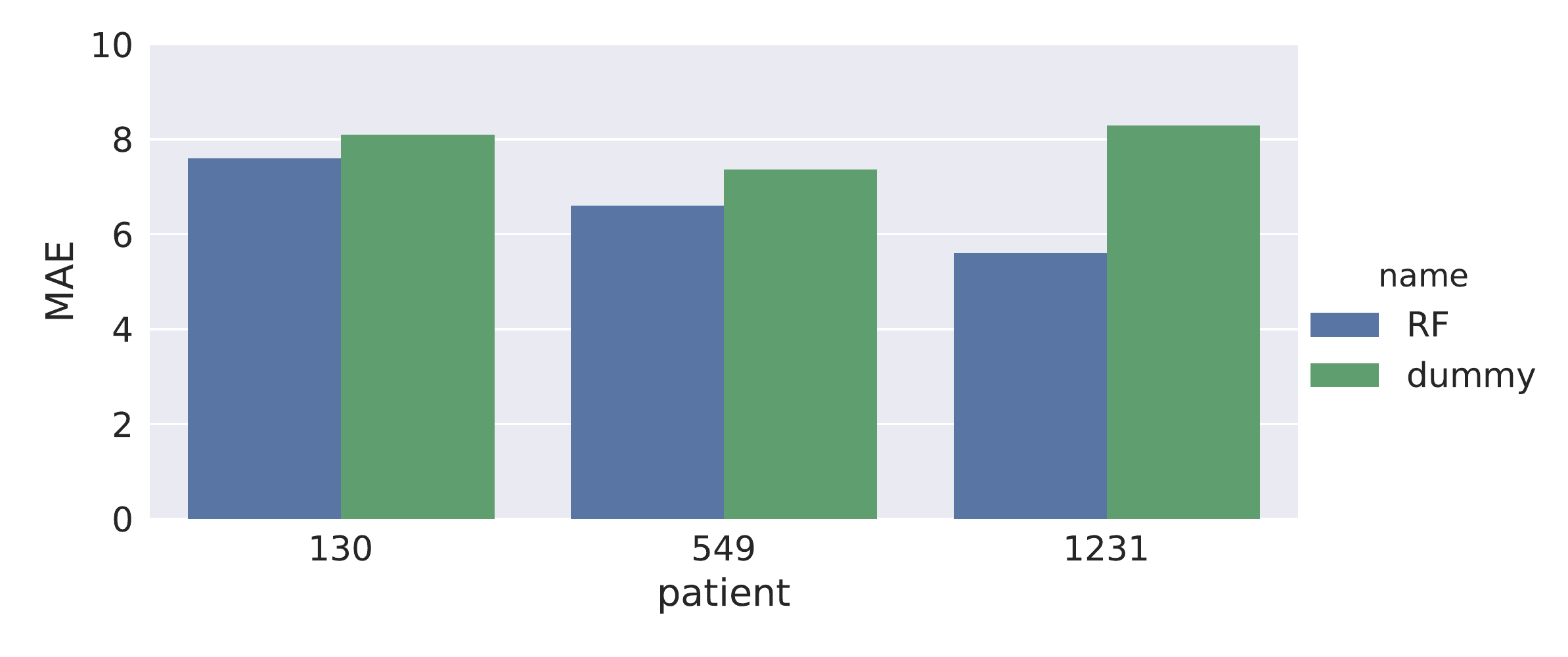} %
    }
  \subcaptionbox{MSE\label{first-subfig}}{
  	\includegraphics[width=.48\columnwidth]{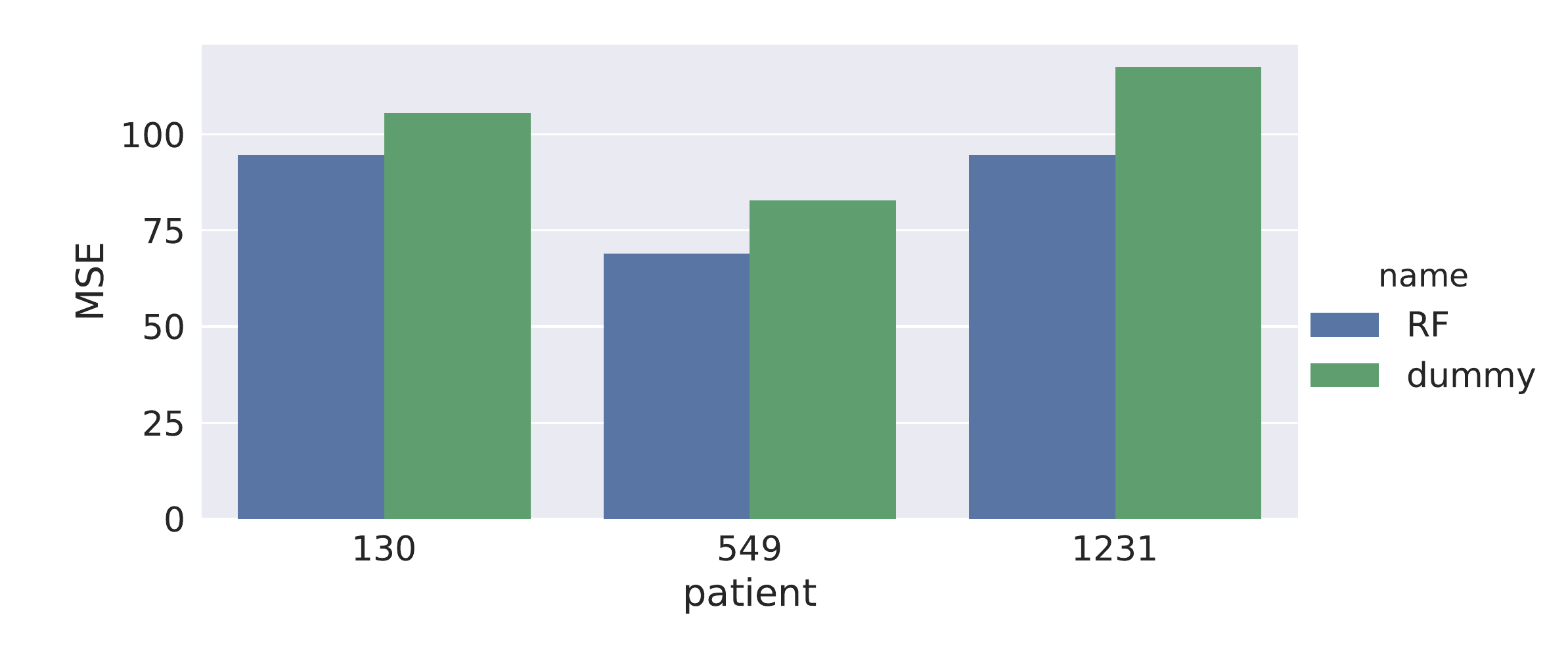}
  }

  \caption{MAE, when trained and tested on different continuous approx 14 day period (a) MAE is is between 6 and 8 which is a decrease in performance seen in Fig.~\ref{fig:offline-results-same}. The performance in terms of MSE has also decreased further.  }

\label{fig:offline-results-diff}
  \end{figure}

If we wish to build a model that can predict longer into the future, this current evaluation is not enough as it was conducted within a continuous approx. 14 days period. Thus we present another experiment where a model is built on one phase (approx. 14 continuous days of data) and the evaluated on another phase (which is a further 14 continuous days over 2 months later).
 In this scenario, seen in Fig.~\ref{fig:offline-results-diff}, the performance significantly decreases to between 6 and 8 MAE, and 65 to 90 MSE. 
 Thus, with this approach, any model for predicting heart rate from acceleration seems to become less accurate with time.
 
\begin{figure} 
   	\includegraphics[width=1\columnwidth]{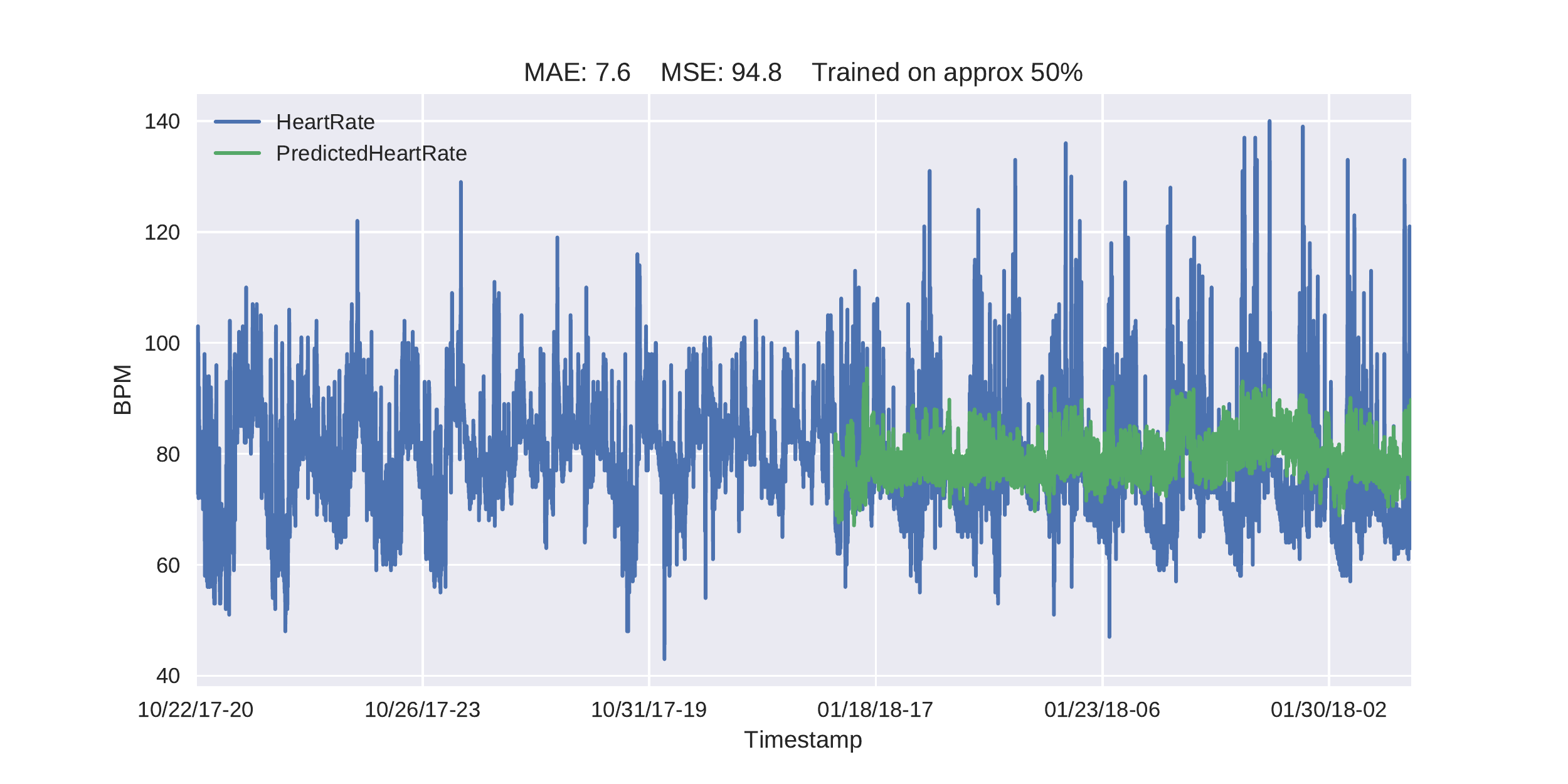}
        \caption{True heart rate versus predicted heart rate using the offline baseline approach for Patient 0130. The left side of the chart was used for training, while the predicted heart rate values are seen on the right hand side.}    \label{hr-pred-offline}
 \end{figure}

 In Fig.~\ref{hr-pred-offline} we can see the actual heart rate plotted against the predicted heart rate for a single patient. 
 It is clear from this that the predictions are not adequate and rarely deviate far from the mean, which is reflected in the MSE (94.8).

\begin{figure}
  \centering
  \subcaptionbox{Patient 0130\label{first-subfig}}{%
 	\includegraphics[width=.3\columnwidth]{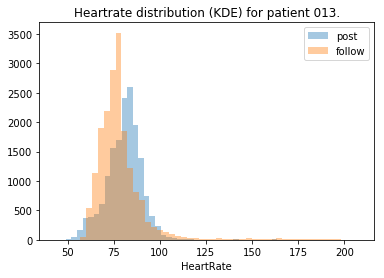} %
  }
  \subcaptionbox{Patient 1231\label{second-subfig}}{%
  	\includegraphics[width=.3\columnwidth]{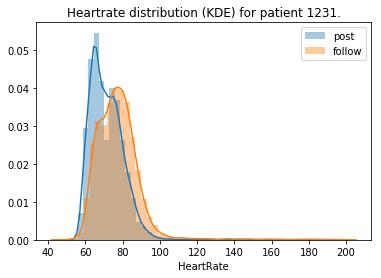}
  }
    \subcaptionbox{Patient 549\label{second-subfig}}{%
  	\includegraphics[width=.3\columnwidth]{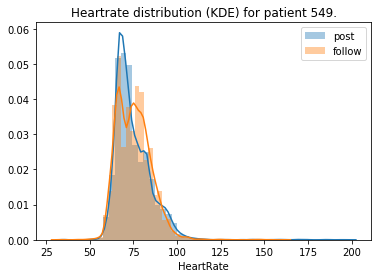}
  }
  	\caption{Heart rate distribution over two different 14 day phases (with around 2 months of time between them) for three patients.}
    \label{hr-dist}
  \end{figure}

 Upon investigation into the distribution of heart rate over these phases, it is apparent why there is such a decrease in performance. In Fig.~\ref{hr-dist} we can see the distribution of heart rate clearly changes over time. In addition, we expect that the concept drift between acceleration and heart rate due to non-acceleration based features, changes frequently within short periods of time.
 Given both of these observations we can model this as a form of \textit{real concept drift}~\cite{Gama:2014:SCD:2597757.2523813} due to the changes of the posterior distribution $P(y|X)$ by developing an approach which more specifically captures the temporal changes in the data between the heart rate ($y$) and acceleration ($X$).

  \subsection{Active Learning}
 If we assume that particular acceleration patterns (e.g. walking) indicate that an increase in heart rate is to be expected, we can reformulate this into an active learning problem where we have access to acceleration on a constant basis, and can query the true heart rate to label the current acceleration sample when required. 
 If we can reduce the number of (energy expensive) heart rate measurements by predicting heart rate using the (energy cheap) acceleration we can increase the battery life of the wearable, which is of significant importance with wearable technology, digital health, Smart Home and Internet of Things (IoT) applications.
  
 Typically, in active learning problems it is possible to query the label (i.e., heart rate) of samples, specifically for data samples for which the label will be particularly useful.
 In the most common of the active learning approaches, it is assumed that there exists a pool of data for which it is possible to choose which data to label. This is however not feasible in our setting, where acceleration data is arriving constantly and we wish to consistently produce a heart rate estimate, and it is not possible to retrospectively measure the heart rate. 
Thus, we must decide at each time point, in an online fashion, if we want to incur the energy cost associated with acquiring the true heart rate or if we can reliably use the acceleration to predict the value. This decision is attenuated by the fact that the $P(y|X)$ may be changing and, therefore, there is a natural temporal aspect which must be taken into account. 
    
 \subsection{Algorithm}
We model these various properties into the procedure depicted in Algorithm~\ref{algo:main}.
The algorithm requires an ensemble of regressors $L$, which provide a measure of uncertainty of the prediction. 
This uncertainty is thresholded with a parameter $O$ which determines the amount of uncertainty tolerated (in terms of the ensemble's variance). 
The algorithm maintains histories of size $N$ to capture the recency.
The previous $N$ prediction variances are stored, with $O$ times the standard deviation of them setting the threshold for determining if the current prediction certainly is an outlier.
This parameter $N$ also controls the initial amount of data to train the ensemble of learners $L$ on, as well as the size of the most recent labeled data to keep for retraining learners in the ensemble.
When there is uncertainty w.r.t. the prediction, the true heart rate is measured and each model in the ensemble is updated with the most recent labeled $N$ samples.
We also take this opportunity to check the actual accuracy of the ensemble, and if any learner has mispredicted by greater than $T$, then they are retrained on the most recent labeled $N$ samples. 
It may be possible for the ensemble to be certain about predictions, but the underlying $P(y|X)$ to have changed without the ensemble being aware.
For this reason, we include a $TTL$ parameter which represents the Time to Live for each learner. After each prediction, their time alive is incremented, and once a learner reaches the $TTL$, it is retrained with the most recent $N$ samples.

  \begin{algorithm}
\SetAlgoLined
  \KwIn{$L$ learners, $N$ size of history, $O$ uncertainty threshold, $T$ error threshold, $TTL$ model time to live}
 Create $L$ learners in ensemble\;
 Train each learner on first $N$ instances\;
 \While{True}{
 $Acc$ = CollectAccDataFor60Seconds()\;
 $HeartRate$, $Variance$ = PredictHeartRate(Acc)\;
  \If{IsOutlier($O$, $Variance$)}{
  	$TrueHeartRate$ = QueryTrueHeartRate()\;
    TeachLearners($Acc$, $TrueHeartRate$)\;
    \For{$l$ $\in$ $L$ }{
    \If{abs($HeartRate$-$TrueHeartRate$) > $T$}{
    	retrain $l$ on most recent $N$ \;
    	}
        }

 }
 \For{ $l$ $\in$ $L$}{
 \If{older than $TTL$}{
     	retrain $l$ on most recent $N$ \;
 	}
    }
      Output $HeartRate$\;

 }
 \caption{PPAW (Predicting Pulse from Acceleration on Wearables)}
 \label{algo:main}
\end{algorithm}

  \subsection{Results}
  
  We apply the presented algorithm to each of the three patients' free living data. We measure the performance of the algorithm using the MAE, which can be interpreted as the the average beats per minute error, the MSE and the percentage of samples for which the true heart rate was measured in order to label the acceleration data sample.
While our algorithm has a number of parameters, due to space constraints we will restrict the investigation to the most influential, the $O$ parameter, which handles uncertainty with the current prediction. 
Here, $N$ is fixed to 5 and $TTL$ to 10.

	\subsection{Varying Uncertainty ($O$)}
\begin{figure}
  \centering
  \subcaptionbox{Patient 0130\label{first-subfig}}{%
      	\includegraphics[width=.3\columnwidth]{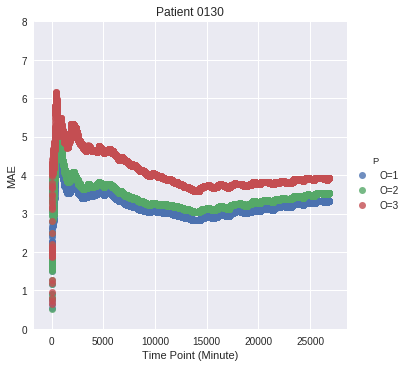}

  }
  \subcaptionbox{Patient 1231\label{second-subfig}}{%
      	\includegraphics[width=.3\columnwidth]{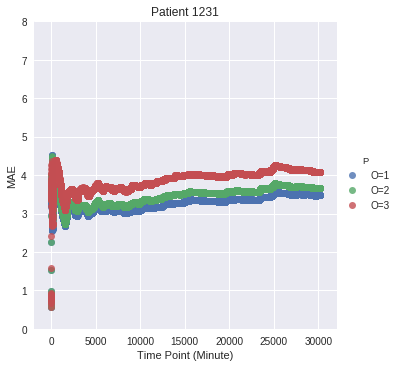}
  }
  \subcaptionbox{Patient 549\label{second-subfig}}{%
      	\includegraphics[width=.3\columnwidth]{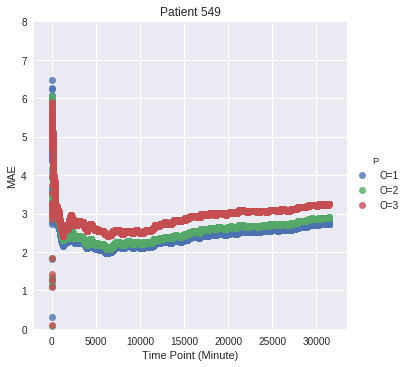}
  }
  \caption{Effect of the $O$ parameter (uncertainty) on MAE.}
  \label{fig:varying0}
\end{figure}

\begin{figure}
  \centering
  \subcaptionbox{Patient 0130\label{first-subfig}}{%
      	\includegraphics[width=.3\columnwidth]{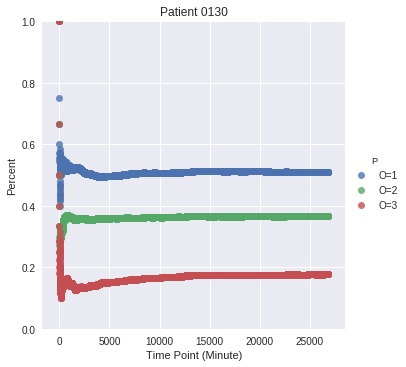}

  }
  \subcaptionbox{Patient 1231\label{second-subfig}}{%
      	\includegraphics[width=.3\columnwidth]{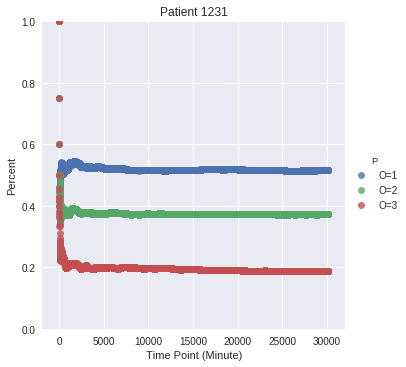}
  }
  \subcaptionbox{Patient 549\label{second-subfig}}{%
      	\includegraphics[width=.3\columnwidth]{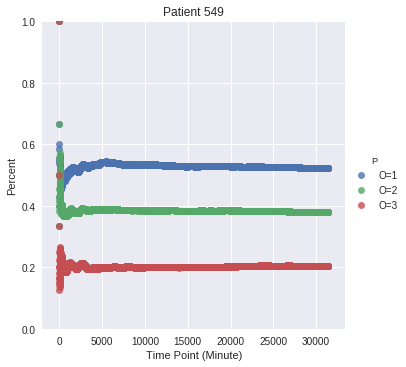}
  }
  \caption{Effect of the $O$ parameter (uncertainty) on the percentage of labels queried. There is a significant decrease in the percentage queried for higher values of $O$, yet only slight decrease in performance for the corresponding $O$ in Fig.~\ref{fig:varying0}.}
  \label{fig:varying0percent}
\end{figure}

	In Fig.~\ref{fig:varying0} we can see the MAE w.r.t. three sensible choices for the uncertainty parameter $O$. While for higher values of $O$ the performance is lower, in all cases the MAE is between 3 and 4. Fig.~\ref{fig:varying0percent} shows the corresponding percentage of labels queried. For higher values of $O$ the number of queries to the heart rate sensor decreased. With a $O$ parameter of 1, the true heart rate was requested around 50\% of the time, while with a value of $O$ the true heart rate was requested around between 15\% and 20\% of the time for all patients. This is a significant decrease in using the heart rate sensor, with only small decrease in performance. In all cases, there is a significant improvement in performance with our online approach compared to the baseline approach.
    
\begin{figure}
  	\includegraphics[width=1\columnwidth]{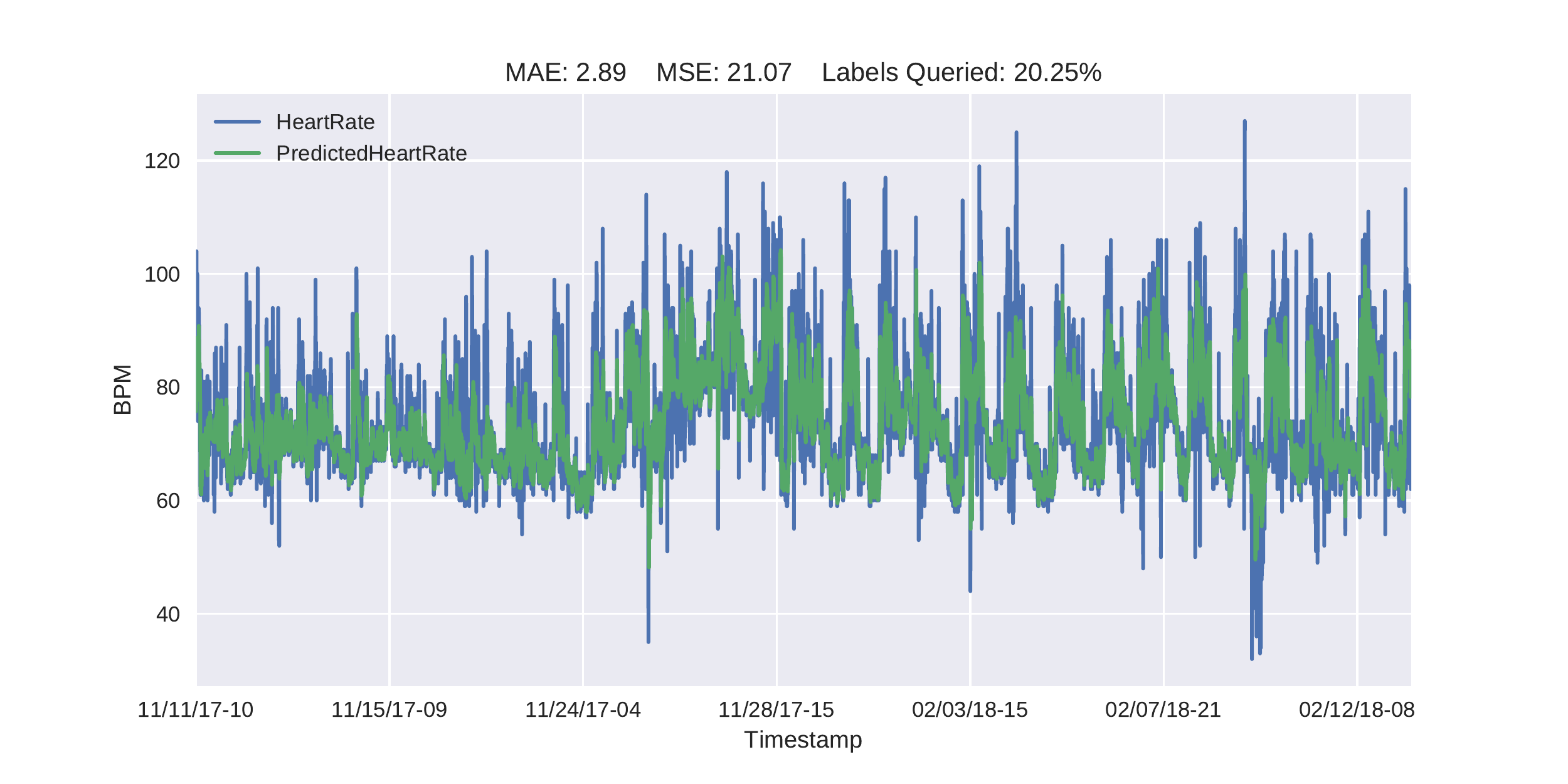}
    \caption{Overview of true heart rate vs predicted heart rate using our approach ($0$=3, $N$=5, $TTL$=10) on patient 549. We can see the predicted heart rate follows the trends exhibited in the true heart rate closely (MAE=2.89 and MSE=21.07) while only requesting measurement of the heart rate 20.25\% of the time.}
    \label{fig:overview-al}
  \end{figure}

    We also plot over all phases the true heart rate (blue) with the predicted heart rate (green) in Fig.~\ref{fig:overview-al}. 
  We can see the significant improvement in our approach over the baseline seen in Fig.~\ref{hr-pred-offline}. The significance is reflected visually in the predictions following the peaks and troughs of the heart rate, which is clearly better in our online approach. Similar results were found for the other patients with significant improvements in MAE and MSE in all cases.

\section{Conclusion}
In this paper we investigate the feasibility of predicting heart rate from acceleration values. A key motivation is to understand the relationship of heart rate to acceleration, and if acceleration can successfully predict heart rate over long periods of time.
In this work we were specifically interested in the changes of heart rate over time in patients recovering from heart valve interventions.
Another motivation is to gain greater energy savings from wearable devices by utilizing the more energy efficient accelerometer to predict heart rate reducing the use of energy-expensive heart rate sensors. We first proposed a baseline method which learns an offline model to predict heart rate from acceleration. However we show that this method performs poorly when attempting to predict the heart rate months later on the same patients. We therefore proposed a method addressing this, as well adapting to non-acceleration based changes to heart rate, using an online active learning approach. We evaluated this method on three patients, each with approximately 4 weeks of free living data, collected over two phases months apart, and show that we can achieve MAEs between 2.5 and 5 heart beats per minute while querying the true heart rate relatively little. In one example the MAE was 2.89 while the heart rate sensor was queried just 20.25\% of the time.

\bibliographystyle{ACM-Reference-Format}
\bibliography{acmart.bib}

\end{document}